\documentclass[journal]{IEEEtran}

\usepackage{tikz} 
\usepackage{amsmath}
\usepackage{url}
\usepackage{xcolor}

\usetikzlibrary{arrows,automata,backgrounds}
\usetikzlibrary{calendar,chains,er,intersections,mindmap,matrix,folding,patterns,plothandlers,plotmarks,shapes,shadings,shadows,scopes,topaths,trees,decorations.pathmorphing,decorations.markings,positioning,fit,petri}

\tikzset{
  block/.style={rectangle, minimum size = 6mm, very thick, draw=blue!50!black!80,top color=blue!50!black!20, bottom color=white, shading angle = 1, rounded corners},
  point/.style={circle, fill=black, scale = 0.6},
  vecArrow_white/.style = {thick, decoration={markings,mark=at position
   1 with {\arrow[semithick]{open triangle 60}}},
   double distance=1.4pt, shorten >= 5.5pt,
   preaction = {decorate},
   postaction = {draw,line width=1.4pt, white,shorten >= 4.5pt}},
  vecArrow_black/.style = {thick, decoration={markings,mark=at position
   1 with {\arrow[semithick,fill=black,line width = 1pt]{triangle 60}}},
   double distance=1.4pt, shorten >= 5.5pt,
   preaction = {decorate},
   postaction = {draw,line width=5pt, black,shorten >= 4.5pt}}
}

\title{Automatic Delay Tuning of a Novel Ring Resonator-based Photonic Beamformer for a Transmit Phased Array Antenna}

\author{Laurens Bliek, Sander Wahls~\IEEEmembership{Senior Member,~IEEE}, Ilka Visscher, Caterina Taddei, Roelof Bernardus Timens, Ruud Oldenbeuving, Chris Roeloffzen~\IEEEmembership{Member,~IEEE}, Michel Verhaegen,~\IEEEmembership{Member,~IEEE}
\thanks{
This research was supported by the Dutch Technology Foundation STW under Project 13336, and partly by the Netherlands Organisation for Scientific Research (NWO) under Project 628.009.012.}
\thanks{
Corresponding author: l.bliek@tudelft.nl.}
\thanks{L. Bliek, S. Wahls and M. Verhaegen are with the Delft
Center for Systems and Control, Delft University of Technology, 
Mekelweg 2, 2628 CD, Delft, Netherlands.}
\thanks{I. Visscher, C. Taddei, R. B. Timens, R. Oldenbeuving and C. Roeloffzen are with LioniX International B.V., P.O. Box 456, 7500 AL,  Enschede, Netherlands.}
}

\def\Ku{\mathrm{K_u}}

\def\OF{1\times 4}
\def\OTF{1\times 24}

\def\degree{^{\circ}}

\begin{document}

\maketitle



\begin{abstract}
We present a novel photonic beamformer for a fully integrated transmit phased array antenna, together with an automatic procedure for tuning the delays in this system.
Such an automatic tuning procedure is required because the large number of actuators makes manual tuning practically impossible.
The antenna system is designed for the purpose of broadband aircraft-satellite communication in the $\mathsf{K_u}$-band to provide satellite Internet connections on board the aircraft.
The goal of the beamformer is to automatically steer the transmit antenna electronically in the direction of the satellite.
This is done using a mix of phase shifters and tunable optical delay lines, which are all integrated on a chip and laid out in a tree structure.
The $\mathsf{K_u}$-band has a bandwidth of $0.5$ GHz.
We show how an optical delay line is automatically configured over this bandwidth, providing a delay of approximately $0.4$ ns.
The tuning algorithm calculates the best actuator voltages based on past measurements.
This is the first time that such an automatic tuning scheme is used on a photonic beamformer for this type of transmit phased array antenna.
We show that the proposed method is able to provide accurate beamforming  ($<11.25\degree$ phase error over the whole bandwidth) for two different delay settings.
\end{abstract}

\thispagestyle{empty}

\section{Introduction}

\begin{figure}[t]
\center
 \begin{tikzpicture}[scale=0.8, transform shape]
     \node [point, label=above:{$\mathsf{Antenna\ elements}$}] at (-3,0) (AE1) {};
     \node [point] at (-2,0) (AE2) {};
     \node [point] at (-1,0) (AE3) {};
     \node [point] at (0,0) (AE4) {};
     \node [point] at (1,0) (AE5) {};
     \node [point] at (2,0) (AE6) {};
     \node [point] at (3,0) (AE7) {};

      \node [circle, scale=8.5,draw=red!60!black!100,densely dotted] at (-3,0) (C1) {};
      \node [circle, scale=7.5,draw=red!50!blue!50!black!100,densely dotted] at (-2,0) (C2) {};
      \node [circle, scale=6.5,draw=blue!60!black!100,densely dotted] at (-1,0) (C3) {};
      \node [circle, scale=5.5,draw=blue!50!yellow!20!black!100,densely dotted] at (0,0) (C4) {};
      \node [circle, scale=4.5,draw=blue!30!yellow!60!black!100,densely dotted] at (1,0) (C5) {};
      \node [circle, scale=3.5,draw=red!20!blue!10!yellow!80!black!100,densely dotted] at (2,0) (C6) {};
      \node [circle, scale=2.5,draw=red!60!yellow!60!black!100,densely dotted] at (3,0) (C7) {};

     \draw [] (35mm,3.5mm)--(-35mm,15.0mm);
     \draw [] (35.8mm,5.5mm)--(-34.2mm,17.0mm);
     \draw [] (36.6mm,7.5mm)--(-33.4mm,19.0mm);
     \draw [] (37.4mm,9.5mm)--(-32.6mm,21.0mm);

		\draw [vecArrow_black] (0,9.5mm)--(3.5mm,21.8mm); 
 \end{tikzpicture}
\caption{Beamforming for a phased array antenna explained. Each antenna element transmits the same signal after a certain time delay. This delay can be chosen in such a way that constructive interference occurs in a certain direction, making it possible to transmit a highly directional signal with a high gain and a focused beam.
	\label{fig:bf}}
\end{figure}
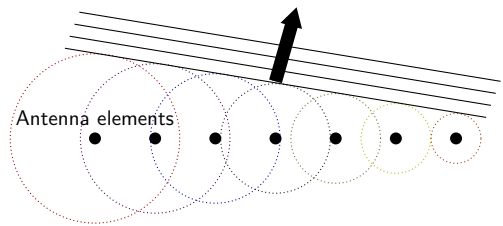

\IEEEPARstart{B}{eamformers} are used to steer the beam of a phased array antenna by controlling the phase added to the signal at each antenna element (AE)~\cite{hansen2009phased}.
This can be done by providing either a phase shift or a time delay to the signal.
When the phase or delay of the signal corresponding to each AE has been set correctly according to the desired beam angle, signals are received or transmitted in the desired direction, while other directions are suppressed.
This results in a highly directional antenna system for which the beam angle can be adapted without any mechanical movement.
See Figure~\ref{fig:bf} for the case of a transmit phased array antenna.
A receive antenna uses the same principle, but in this work we only consider a transmit antenna.

\begin{figure*}[tb]
\centering
\includegraphics[width=0.8\textwidth]{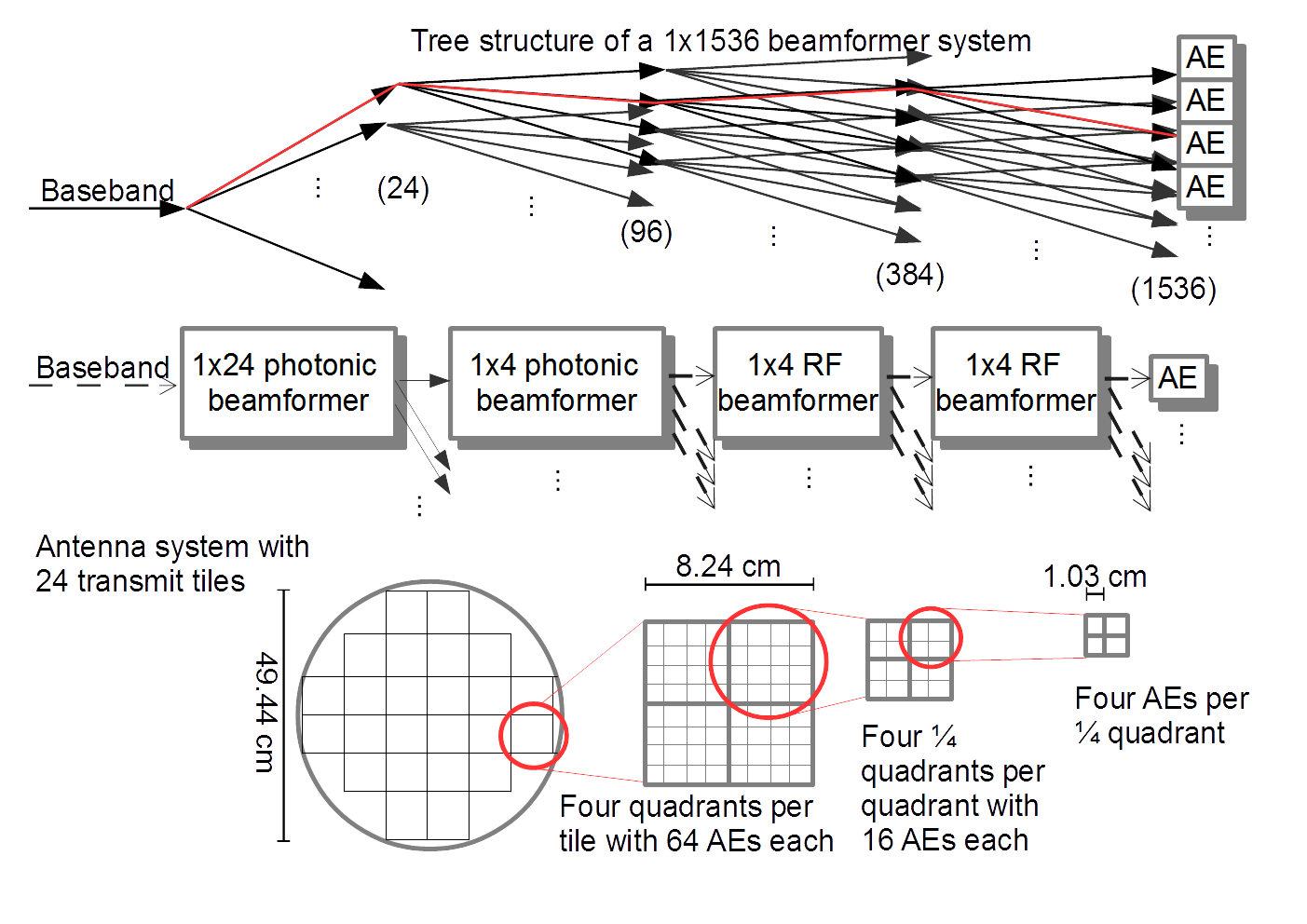}
\caption{From top to bottom: tree structure of the full beamformer system with $1536$ antenna elements (the highlighted path is explained in the remainder of the figure). 
Photonic and RF beamformer systems, where dashed arrows indicate signals in RF domain and solid arrows indicate signals in the optical domain (the photonic beamformers are shown in more detail in Figure~\ref{fig:OBFNschematic}).
Full antenna system architecture with $24$ transmit tiles containing $64$ antenna elements each. 
}
 \label{fig:PAAschematic}
\end{figure*}

In order to get the same signal to each AE with a certain delay, a beamformer for a transmit phased array antenna consists of splitters and delay elements.
These can be arranged in a tree structure to reduce the required number of delay elements~\cite{novel1,novel2,zhuang2010ring}.
The beamformer considered in this work is part of a transmit phased array antenna system with $1536$ AEs.
It uses a tree structure with four different splitting stages, as shown on the top of Figure~\ref{fig:PAAschematic}.
This system is designed to be used for aircraft-satellite communication in order to satisfy the ever-increasing demand of high-speed Internet on board of aircrafts.
This large number of AEs is required to make sure that the signals have enough power when they arrive at the satellite~\cite{verpoorte2010architectures, van2014design}.
The system establishes an uplink with the satellite in the $\Ku$-band ($14.0$ to $14.5$ GHz frequency range). 
This is the range used to provide satellite Internet connections on board the aircraft.
Therefore, the system will operate under a bandwidth of $0.5$ GHz.

\begin{figure*}[tb]
\centering
\vspace{0mm}\includegraphics[width=0.9\textwidth]{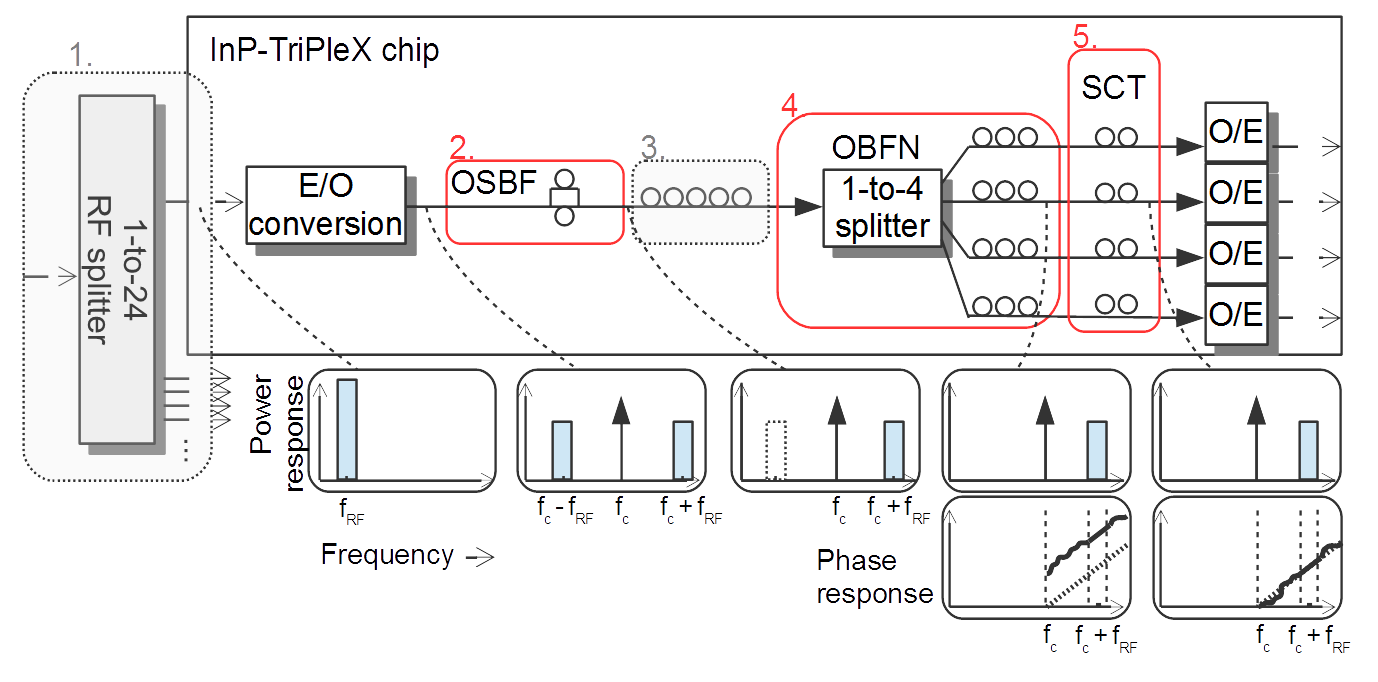}
\caption{Photonic beamformer design with several highlighted subsystems. Subsystems $1.$ and $3.$ belong to the $\OTF$ photonic beamformer in Figure~\ref{fig:PAAschematic}, while the other three subsystems belong to the $\OF$ photonic beamformer. Dashed arrows indicate signals in RF domain and solid arrows indicate signals in the optical domain. The small circles indicate the presence of on-chip optical ring resonators. Power and phase responses of the signal at several locations on the chip are shown on the bottom, where the desired phase response is indicated by a dashed line.
}
\label{fig:OBFNschematic}
\end{figure*}

The delay that needs to be provided for one path of the tree structure depends on the distance between the AEs, which has been chosen to be $1.03$ cm in this work as explained later.
AEs that are close together correspond to only a small delay difference. 
This is why the middle of Figure~\ref{fig:PAAschematic} contains different types of beamformers: radio frequency (RF) beamformers are better suited for very small delay values, while photonic beamformers (explained in the next section) are better suited for larger delay values~\cite{novel1}.
The bottom of Figure~\ref{fig:PAAschematic} shows a top view of how the AEs are situated on the full antenna system that consists of $24$ transmit tiles.
The transmit tiles are $8.24$ by $8.24$ cm in size and contain $8\times 8$ AEs each, resulting in a $1\times 1536$ transmit scheme with $1536$ AEs in total.
Both the photonic and RF beamformers are integrated in the transmit tile, giving rise to a modular system.

The antenna system is an adapted version of an earlier proposed phased array receiver~\cite{van2014design} that is to be used in transmit mode.
In order to tune the beamformer actuators of these earlier phased array antenna systems, either a manual tuning procedure was used~\cite{novel1, novel2}, which is only possible for a small number of actuators, or a nonlinear optimization algorithm based on a physical model of the system was used~\cite[Sec. 6]{zhuang2010ring}.
However, the approach that uses a nonlinear optimization algorithm is very sensitive to model errors~\cite{ifacpaper}, requires a labor-intensive measurement procedure for certain parts of the physical model~\cite[App. B]{zhuang2010ring}, and can not be used while the system is already running.
In this work we describe the beamformer for the transmit phased array antenna from Figure~\ref{fig:PAAschematic} together with its requirements, and we show how to tune this system with a different optimization algorithm.
This online algorithm is not sensitive to model errors and can be used while the system is running.

\section{Fully Integrated Transmit Phased Array Antenna}

\subsection{Beamformer requirements}

In order to determine the requirements of the beamformer, the delay between the AEs needs to be calculated.
This depends on two factors: the angle of the beam, and the distance between the AEs.
For rectangular array grids, the latter should be equal to half the wavelength corresponding to the highest frequency used in the application~\cite[Sec. V]{novel1}, which is $14.5$ GHz in this case.
This leads to a distance of $1.03$ cm between two consecutive AEs.
As for the beam angle, this is chosen to be no more than $60$ degrees compared to the normal (i.e. a beam sent straight upwards).
The maximum delay $\tau_{\mathrm{max}}$ that needs to be provided by each beamformer in Figure~\ref{fig:PAAschematic} can now be calculated as follows:
\begin{align}
	\tau_{\mathrm{max}} = \sin(60\degree) d_{\mathrm{max}}/c_0,
\end{align}
where $d_{\mathrm{max}}$ is the maximum distance between the centers of the different elements of the beamformer and $c_0$ is the speed of light.
For the rightmost RF beamformer in Figure~\ref{fig:PAAschematic}, $d_{\mathrm{max}} = \sqrt{2}\cdot 0.0103$ m (distance between the centers of two diagonally intersecting AEs), for the other RF beamformer $d_{\mathrm{max}} = \sqrt{2}\cdot 2 \cdot 0.0103$ m, for the $\OF$ photonic beamformer $d_{\mathrm{max}} = \sqrt{2}\cdot 4 \cdot 0.0103$ m, and for the $\OTF$ photonic beamformer $d_{\mathrm{max}} = \sqrt{5^2+1^2}\cdot 0.0824$ m.
This leads to the following maximum delays for each beamformer in Figure~\ref{fig:PAAschematic}:
$\tau_{\mathrm{max}} = 42$ ps for the rightmost RF beamformer, $\tau_{\mathrm{max}} = 84$ ps for the other RF beamformer, $\tau_{\mathrm{max}} = 168$ ps for the $\OF$ photonic beamformer, and $\tau_{\mathrm{max}} = 1.214$ ns for the other photonic beamformer.

After calculating the maximum delays for each beamformer subsystem, the requirements for the phase can be calculated.
The phase response of each beamformer should be a linear function of the frequency, with a slope equal to $-2\pi \tau$, where $\tau$ is the required delay.
Traditionally, phase shifters are used for beamformer subsystems, which provide a phase shift that is a constant function of the frequency.
In other words, phase shifters provide a linear phase response with a slope of $0$.
For very small delays (e.g. $\tau_{\mathrm{max}} < 100$ ps in this case), using such a constant phase response with a slope of $0$ instead of $-2\pi \tau$ gives a negligible slope mismatch, but for larger delays this mismatch will have a negative influence on the antenna beam.
This problem is called beam squint.

The effects of beam squint are negligible in the two RF beamformer subsystems in Figure~\ref{fig:PAAschematic}.
Therefore, these beamformers are based on phase shifters that change the phase of the RF signal.
The photonic beamformers, however, need to deal with larger delays.
Therefore, they make use of optical delay lines rather than phase shifters.
The key principle here is that the RF signals are converted to the optical domain, where their phase is adjusted accordingly, and then they are converted back to RF domain.
This combination of RF and photonic signal processing has many advantages such as low loss, low weight, and large bandwidth, especially when implemented on a chip~\cite{novel1, novel2, capmany2007microwave, shi2016experimental, zhang2017photonic},~\cite{marpaung2013integrated}.
In this work, each transmit tile from Figure~\ref{fig:PAAschematic} is connected to a corresponding chip, as explained later on in this section.

The way the photonic beamformers provide a linear phase response with the correct slope is by using a cascade of optical ring resonators as tunable optical delay lines, arranged in a tree structure~\cite{novel1, novel2, zhuang2006single}.
These ring resonators control the phase response of the signal by using thermo-optic heater actuators.
Though some fluctuations in the phase response are allowed, these fluctuations should not be larger than $11.25\degree$~\cite{novel2}.

\subsection{Photonic beamformer chip design}

Figure~\ref{fig:OBFNschematic} shows the chip design of the photonic beamformers.
The chip contains optical ring resonators that each belong to different subsystems.
The effect of the different subsystems on the frequency response of the signal is shown as well.
An RF splitter (subsystem $1$. in Figure~\ref{fig:OBFNschematic}) makes sure that the same baseband signal arrives at all of the $24$ transmit tiles shown in Figure~\ref{fig:PAAschematic}.
Each transmit tile is connected to one chip containing photonic and RF beamformer systems.
The photonic beamformers use a combination of TriPleX\texttrademark{} waveguide technology~\cite{roeloffzen2018low} 
and indium phosphide (InP). 
TriPleX\texttrademark{} is a silicon nitride planar waveguide technology developed by LioniX International.

The corresponding power response of the electrical signal entering the chip in Figure~\ref{fig:OBFNschematic} is that of a broadband signal centered around the frequency $f_{RF} = 14.25$ GHz in the $\Ku$-band.
This signal is modulated on an optical carrier with a Mach-Zehnder modulator (electrical-to-optical conversion), after which the optical signal consists of the optical carrier $f_c$ in the THz range and two sidebands.
One of these sidebands is suppressed by the optical sideband filter (OSBF, subsystem 2. in Figure~\ref{fig:OBFNschematic}).
This is done to reduce the required bandwidth\cite{novel1}.
After this stage the signal goes through a cascade of five ring resonators (subsystem $3$. in Figure~\ref{fig:OBFNschematic}) that 
supply the required delay for one branch of the $\OTF$ beamformer.
As mentioned earlier, the maximum delay that needs to be provided by the $\OTF$ beamformer is $\tau_{\mathrm{max}}=1.214$ ns.
This should be no problem for a cascade of five ring resonators, as beamforming with a cascade of only three ring resonators has been illustrated for delays of more than $1.2$ ns with a bandwidth of $0.5$ GHz using a manual tuning approach~\cite{orr1}. 
With five ring resonators, these results can be achieved with even higher accuracy.

The signal then arrives at a $\OF$ optical beamforming network (OBFN, susbystem 4. in Figure~\ref{fig:OBFNschematic}), consisting of an optical splitter and more ring resonators.
These ring resonators provide a linear phase response for a maximum delay of $168$ ps, as calculated earlier in this section.
As seen in the phase response in Figure~\ref{fig:OBFNschematic}, the linear phase response is only provided in the region of interest, the sideband around frequency $f_c + f_{RF}$.
However, the OBFN only makes sure that this linear phase response has the correct slope, not the correct absolute phase.
The last subsystem (subsystem $5$. in Figure~\ref{fig:OBFNschematic}) provides an additional phase shift so that the correct phase response is provided at the sideband and also at the optical carrier frequency $f_c$.
This is again done with optical ring resonators, using a principle called separate carrier tuning (SCT)~\cite{SCTpaper} (see also Section~\ref{sec:SCT}).
Finally, the signal is converted back to RF domain with photodetectors (optical-to-electrical conversion), after which it can be converted to the correct frequency range for transmission ($14.0$ to $14.5$ GHz).

The ring resonators in the beamformer subsystems are all actuated with two heaters: one for the phase, and one for the tunable coupling~\cite{stokes1982all, orr1}. 
By changing the voltages of these two heater actuators for several ring resonators, the magnitude and phase response of the subsystems can be altered. 
There are $27$ ring resonators shown in Figure~\ref{fig:OBFNschematic}, resulting in $54$ heaters.
Three other heaters in the OSBF subsystem and three heaters in the $1$-to-$4$ splitter are not shown in the figure but increase the number of heaters to $60$.
So the total number of heaters in the two photonic beamformer stages is $24\times 60 = 1440$.
One of the main reasons for investigating automatic procedures is that tuning all these heaters correctly by hand would not be practical.

\section{Automatic Tuning Results}

\subsection{Automatic tuning method}

In this paper we look at an automatic tuning procedure for one branch of the OBFN subsystem.
The traditional approach to automatically tune one branch of the photonic beamformer is to use a nonlinear optimization algorithm that minimizes a performance metric~\cite[Sec. 6]{zhuang2010ring}.
An example performance metric is the mean square error (MSE) between the desired and actual phase response of the system ($\phi_{D}(f,V)$ and $\phi_{P}(f,V)$ respectively):
\begin{align}\label{eq:physical}
\min_V \sum_{k}\left(\phi_{D}(f_k,V)-\phi_{P}(f_k,V)\right)^2.
\end{align}
Here, $V$ represents a vector of actuator voltages while the $f_k$ represent the different frequencies that are relevant to the system.
The actual phase response $\phi_{P}(f,V)$ can be calculated from physical models of the heaters and optical ring resonators.

The procedure described above has several disadvantages.
First of all, we have shown in earlier work that even very small errors in the physical model can have a large detrimental effect on the accuracy of the procedure~\cite{ifacpaper}.
Such model errors can never completely be avoided.
Second, although physical models are available for each system component, the heater actuators influence each other by means of electrical and thermal crosstalk.
This crosstalk can also be modeled, but this requires a number of measurements that is at least equal to the square of the number of heater actuators~\cite[App. B]{zhuang2010ring}.
This becomes a problem when the number of heater actuators is large, and it gives even more room for model errors.
Finally, the procedure has to be performed before actually running the system, and any changes to the system are not automatically taken into account.

To avoid the drawbacks mentioned above,
recently automatic tuning algorithms based on machine learning techniques have been derived~\cite{ifacpaper, DONEpaper, sitbpaper}.
These algorithms use system measurements in the performance metric.
Instead of calculating the phase response from physical models, the phase response is measured directly:
\begin{align}\label{eq:databased}
\min_V \sum_{k}\left(\phi_{D}(f_k,V)-\phi_{M}(f_k,V)\right)^2.
\end{align}
Here, $\phi_{M}(f,V)$ denotes the measured phase response.
This data-based algorithm learns the relation between the actuator voltages and the performance metric in~\eqref{eq:databased} and chooses those voltages that optimize this metric.
After several iterations, the algorithm converges to a setting of actuator voltages $V$ with the best performance according to the data-based performance metric~\eqref{eq:databased}.
Here, each iteration consists of updating the performance metric with the measured phase response, finding the optimal values for $V$, and setting the actuator voltages to these values and performing a new measurement.
\begin{figure}[tb]
\centering
\includegraphics[width=0.7\columnwidth]{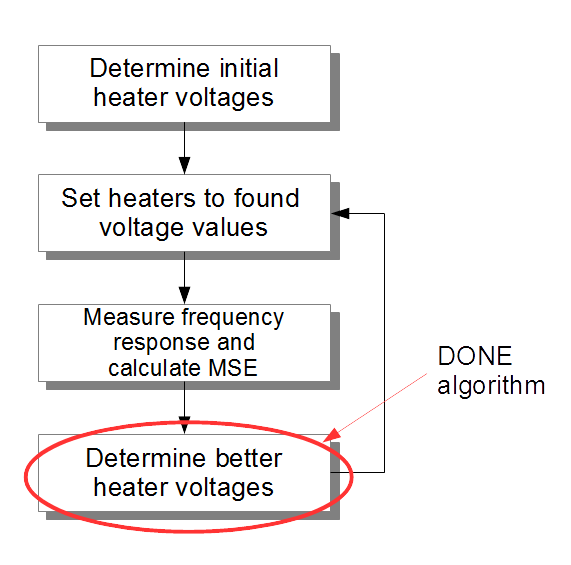}
\caption{Schematic for the automatic tuning method.}
\label{fig:doneschematic}
\end{figure}
In this work this is done with the data-based online nonlinear extremum-seeker (DONE) algorithm~\cite{DONEpaper}.
See Figure~\ref{fig:doneschematic}.
This algorithm is especially designed for systems where the measurements take some time to be performed, where measurement noise is present, and where the relation between the MSE and the actuators is too complicated to model accurately.
The beamformer system described in this paper is an example of such a system, and the DONE algorithm has already been successfully applied to a simulation of a beamformer system with $8$ antenna elements~\cite{DONEpaper}.
Using such an automatic tuning procedure on this system with over a thousand actuators is unavoidable if it is ever to be used in a real life application.
\emph{Besides this, the DONE algorithm only uses the MSE in~\eqref{eq:databased}, which is a single scalar value.}
In a practical situation, one can thus use very simple metrics such as, e.g., the total output power in order to guide the automatic tuning algorithm. In this paper, where the performance is shown of both the automatic tuning method and a human expert that relies on the complete phase response for tuning, we however used the average squared phase error in order to guide the automatic tuning method such that both use a similar criterion. We emphasize that the human expert knew the phase error \emph{	per frequency} (by looking at the frequency response), while the automatic tuning method only knew the mean squared phase error that is obtained \emph{after averaging over all frequencies}.
It should be noted that
the traditional approach with the nonlinear optimization method also uses a scalar-valued objective, but it becomes very slow and struggles with local minima when the output power is used as the objective~\cite[Sec. 4.4.2]{blokpoelthesis}.

Unlike the traditional approach~\eqref{eq:physical}, the data-based approach~\eqref{eq:databased} does not require a  physical model (including a model for the heater crosstalk) and is therefore not sensitive to model errors.
It is also an online algorithm, which means that it can be used while the system is running, and it will continually search for better configurations.
This can also easily be adapted to allow the procedure to work for systems that change over time, which is crucial when applied to antenna systems on moving vehicles such as aircrafts.
On moving vehicles, the beam angle changes over time, and therefore the objective function also changes over time.

\subsection{Optical sideband filter tuning}

Earlier in this paper, several subsystems of the photonic beamformer were described.
See Figure~\ref{fig:OBFNschematic}.
The OSBF and SCT subsystems drastically reduce the required bandwidth of the system~\cite{novel1, SCTpaper}. 
The goal of the OSBF is to filter out one of the sidebands resulting from the modulation scheme used for the E/O conversion, as shown in the frequency responses of Figure~\ref{fig:OBFNschematic}.
The result is an optical single sideband full carrier modulation that can be used by the OBFN.
With one sideband filtered out, beamforming would need to be performed over the sideband, the optical carrier frequency, and the frequencies in between, giving a total of $14.5$ GHz instead of the whole region of $29$ GHz.
To further reduce the bandwidth, the SCT subsystem is used to ensure that the correct phase is achieved at the optical carrier frequency (see Section~\ref{sec:SCT}). 
If both the OSBF and SCT subsystems are tuned correctly, a linear phase response only needs to be provided in one sideband with a bandwidth of $0.5$ GHz.

\begin{figure}[tb]
\centering
\includegraphics[width=0.7\columnwidth]{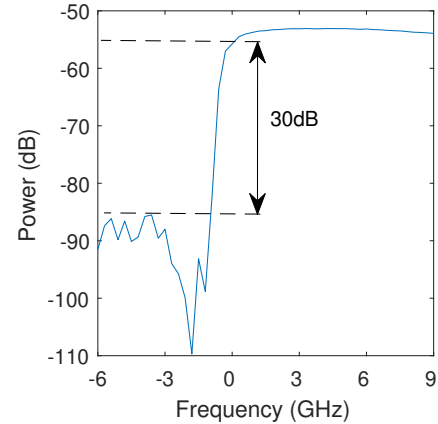}
\caption{Measured power response of the full beamformer system after tuning the optical sideband filter by hand.
Frequencies are shown relative to the optical carrier frequency.
}
\label{fig:OSBFtune}
\end{figure}

Figure~\ref{fig:OSBFtune} shows the power response of the full beamformer system after tuning the OSBF subsystem by hand. 
This was done by measuring the power response with a vector network analyzer (VNA) and tuning the OSBF heaters until the desired response was achieved.
The shown frequencies are relative to the optical carrier frequency.
The response features a stopband, that filters out one of the sidebands as shown in Figure~\ref{fig:OBFNschematic}, and a passband that keeps the other sideband.
The difference between the stopband and passband is 30 dB.

The automatic tuning method has also been applied to the OSBF subsystem, see the end of Section~\ref{sec:autoOBFN}.
No matter which tuning method is used, once tuned, the heaters can be set to the desired values within $200 \ \mathrm{\mu s}$.

\subsection{Automatic optical beamforming network tuning}\label{sec:autoOBFN}

The optical beamforming network (OBFN) subsystem is used to provide different delays to the signals of each antenna element in such a way that a strong signal is transmitted in one specific direction by positive interference.
The correct delays can be calculated from the desired beam angle and the distance between two antenna elements, as shown in the introduction.
This leads to a maximum delay of $168$ ps for the $\OF$ photonic beamformer, which is the one investigated in this work.
Larger delay values were considered in this work to show that the maximum delays can indeed be achieved, to allow different antenna configurations or scan angles, and for better visibility of the measurements.

\begin{figure*}[tb]
\centering
\includegraphics[width=0.9\textwidth]{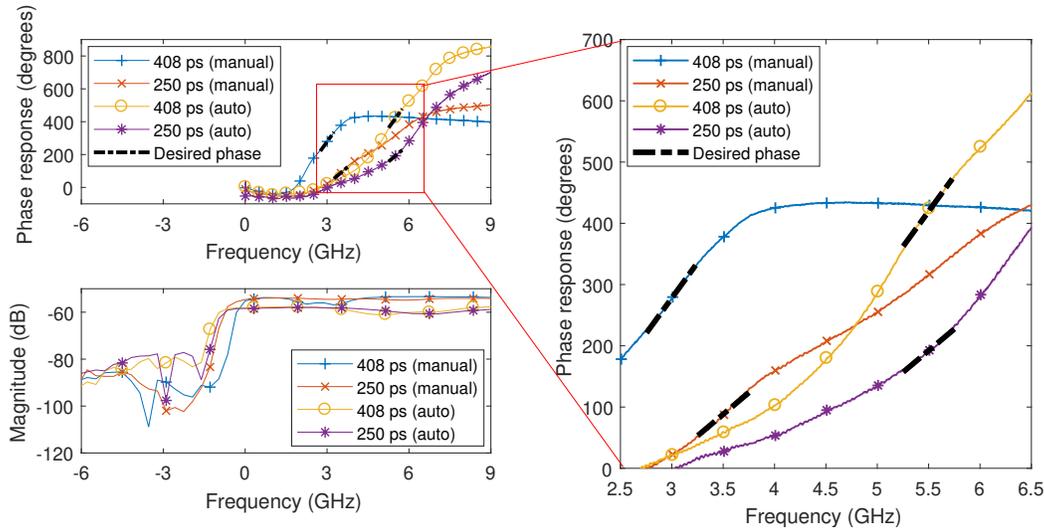}
\caption{Measured phase response (top) and power response (bottom) of one path of the full beamformer system from Figure~\ref{fig:OBFNschematic} after tuning the optical beamforming network automatically (using the DONE algorithm) and by hand for two different delay settings. The right side zooms in on the highlighted region.
Desired phase responses are shown with dashed lines.
The maximum difference between the measured and desired phase response for the four plots is shown in Table~\ref{tab:ripple}.
Frequencies are shown relative to the optical carrier frequency, and the phase responses are shown relative to the phase response of the system with no delay.
}
\label{fig:OBFNtune}
\end{figure*}

The desired group delay response for each OBFN path is a flat response with a value between $0$ ps and the maximum group delay value, depending on the desired beam angle.
The group delay response can be made flat by using a serial configuration of ring resonators for each OBFN path~\cite{orr1}.
This flat response has to be provided over the whole frequency range of interest, which would require a combination of many ring resonators.
However, by implementing the SCT scheme~\cite{SCTpaper} explained in Section~\ref{sec:SCT}, the only part of the frequency domain for which a flat group delay response needs to be provided is the sideband that is not filtered by the OSBF, as shown in Fig.~\ref{fig:OBFNschematic}.
This greatly reduces the required number of ring resonators and therefore the complexity of the system, because there is a trade-off between the required number of ring resonators, the maximum group delay, and the bandwidth~\cite{novel1}.

The procedure for tuning the OBFN is similar to the one for tuning the OSBF: the group delay response is measured with a VNA, and the heaters of each ring resonator are tuned until the desired group delay response is achieved.
Here, we made use of the automatic tuning method described in this paper.
The top of Figure~\ref{fig:OBFNtune} shows the resulting phase response of both the manual and the automatic tuning procedures of one path of the photonic beamformer system for two different delay settings ($250$ ps and $408$ ps).
The $1$-to-$4$ splitter shown in Figure~\ref{fig:OBFNschematic} has been set to provide this path with $100\%$ of the signal power.
The manual tuning procedure aimed at providing a linear phase response for a bandwidth of at least $0.5$ GHz, though no exact frequencies were provided.
The automatic procedure aimed at providing a linear phase response for the same bandwidth, but around a frequency of $5.5$ GHz.
Since all signals can be converted to the correct transmit frequencies later (for example with a RF mixer), only the bandwidth is considered relevant in this work, not the exact frequencies.
All results were time averaged to reduce noise.

\begin{table}[b]
\caption{Maximum phase error in degrees for tuning the OBFN subsystem automatically (A) and by hand (M) for different delay values. All errors are well below the maximum allowed error of $11.25\degree$. \label{tab:ripple}}
\centering
\begin{tabular}
{|p{0.20\columnwidth}|c|c|c|c|}
\hline
Delay & $408$ ps & $250$ ps & $408$ ps & $250$ ps \\
\hline
Method & (M) & (M) & (A) & (A)\\
\hline
Max error & $\mathbf{6.2\degree}$ & $\mathbf{4.9\degree}$ & $\mathbf{5.5\degree}$ & $\mathbf{6.4\degree}$\\
\hline
\end{tabular}
\end{table}

The difference between the desired and the measured phase response must remain within $11.25\degree$ in the bandwidth of interest.
The maximum errors are shown in Table~\ref{tab:ripple}.
As can be seen, all phase errors remain well within the requirements.
This shows that the automatic tuning procedure can be used instead of the manual tuning procedure for the photonic beamformer system.

The bottom of Figure~\ref{fig:OBFNtune} shows the corresponding power response.
For the manual tuning method this is similar to the one shown in Figure~\ref{fig:OSBFtune}, with some influence of the ring resonators.
For the automatic tuning method, the magnitude response was tuned automatically too, using the same procedure for the OSBF as for tuning the OBFN, with the magnitude instead of the phase response.
However, some improvement is possible here, as the loss in the passband could still be reduced.
It seems that the MSE criterion does not work too well with the logarithmic scale.
The MSE criterion is especially sensitive to data points in the stop band in this case.
This was already somewhat circumvented by scaling the data points in the stopband by a factor $0.3$, but other objective functions should be considered in the future.

After tuning the OSBF, while tuning the OBFN, heater crosstalk could cause an unwanted perturbation in the OSBF filter response.
For this reason, the OSBF actuators were still included in the tuning procedure while tuning the OBFN automatically, allowing the algorithm to counteract the heater crosstalk and fine-tune the system.
This lead to stability in the filter response of the OSBF, i.e., there was no deterioration in performance in the OSBF while tuning the OBFN.

\subsection{Separate carrier tuning}\label{sec:SCT}

\begin{figure*}[tb]
\centering
\includegraphics[width=0.85\textwidth]{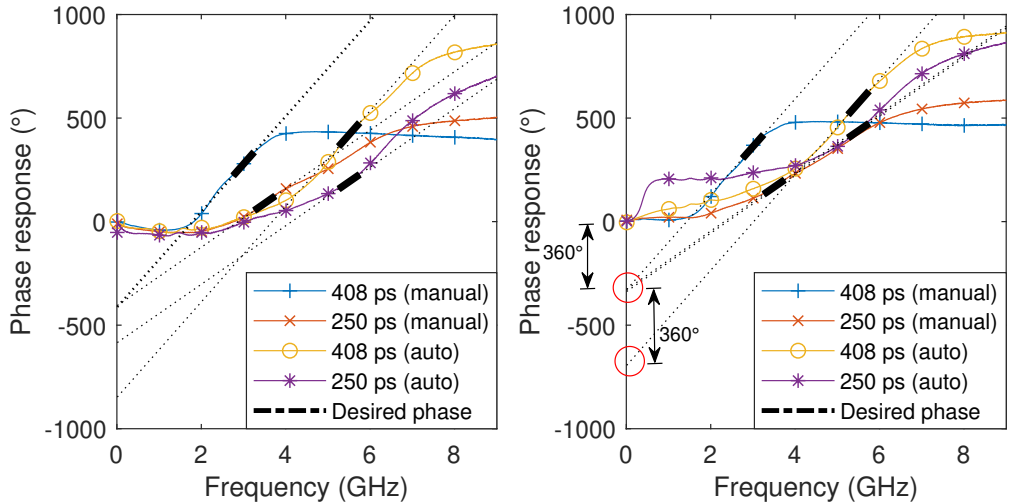}
\caption{Measured phase response of the full beamformer system before (left) and after (right) manually tuning the SCT subsystem. The word `auto' in the legend refers to the automatic tuning of the OSBF and OBFN subsystems. Using the SCT principle, the height of the phase responses is adapted in such a way that the corresponding linear responses meet at the optical carrier frequency, up to multiples of $360$ degrees, as indicated by the red circles. These results have also been published in~\cite{Roeloffzen18}.}
\label{fig:SCT}
\end{figure*}

The results of the previous subsection show that both the manual and automatic tuning procedures result in a phase response with the correct slope, satisfying the requirements.
However, besides the slope of the phase response, the height of the phase response also needs to be set correctly~\cite{SCTpaper, Roeloffzen18}.
Where the slope of the phase response is controlled by the OBFN subsystem, the height of the phase response is controlled by the separate carrier tuning (SCT) subsystem.
This is also shown in Figure~\ref{fig:OBFNschematic}.
More information about the SCT subsystem can be found in~\cite{SCTpaper}.

Like the other subsystems, the SCT subsystem makes use of optical ring resonators.
Two ring resonators are tuned in such a way that they affect the phase response in the region between the optical carrier frequency and the sideband.
This region contains no information, so it is not necessary to follow the desired linear phase response here, which would require a large quantity of ring resonators because of the large difference in frequency between the optical carrier and the sideband.
Tuning the two ring resonators of the SCT subsystem gives an extra phase shift while leaving the shape and the slope of the phase response unaffected.
This ensures that the phase response follows the desired linear phase response at the optical carrier frequency too.
It should be noted that phase differences of multiples of $360\degree$ do not affect the system~\cite{SCTpaper}.
This makes it theoretically possible to use only one ring resonator for this adjustment, though in practice it is easier to use two of them.
The ring resonators in the SCT subsystem were adjusted by hand after tuning their heater actuators in such a way that the ring resonators only operate in the region near the optical carrier frequency.

Figure~\ref{fig:SCT} shows the effect of not using the SCT subsystem on the left.
Although the phase response has the desired slope, there is a phase mismatch at the optical carrier frequency (at $0$ GHz).
Tuning the ring resonators of the SCT subsystem solves this problem.
The effect of tuning the SCT subsystem is shown in the same figure on the right.
This time, the phase responses have been shifted up or down in such a way that the corresponding linear phase responses are a multiple of $360$ degrees off at the optical carrier frequency.
The accuracy can still be improved, however: due to their proximity on the photonic chip, the heaters of the SCT subsystem slightly influence the heaters of the OBFN subsystem, which leads to small changes to the slope of the phase response.
This causes the desired phase response at the carrier to be about $30$ degrees off, so exact multiples of $360$ degrees are not yet achieved.
These errors are unacceptable for practical applications.
However, using the same automatic tuning procedure for this subsystem together with the other two subsystems could circumvent this problem.
This remains for future work. 

\subsection{Towards a fully automated transmit antenna tile}\label{sec:towards}

Besides testing the proposed automatic tuning procedure on all of the three different subsystems described in this section and improving the objective for the OSBF subsystem, one more step is required to achieve a fully automated transmit antenna tile in a practical situation. First of all, since phase response measurements require a measurement set-up that is too cumbersome for a real-life application, the effect of using a different criterion needs to be investigated. As mentioned earlier, the proposed automatic tuning procedure uses only a scalar value to tune the system, even with the phase measurements used in this work, therefore the criterion used in practice is allowed to be scalar-valued. The output power of the system is an example of such a more practical criterion. Traditional nonlinear optimization algorithms based on physical models of this criterion are known to struggle with local minima~\cite[p. 43]{blokpoelthesis}. However, we found that using the DONE algorithm on this same criterion~\cite[eq. 4.4]{blokpoelthesis} in a simple simulation, the global optimum was found, even when tuning four beamformer branches at the same time. The phase errors were also within $11.25\degree$ for all four branches. This indicates that it should be possible to tune all paths of a $\OF$ photonic beamformer using the automatic tuning procedure with a power criterion. A next step would be to verify this experimentally in a real system.

The second step concerns the tuning of the $\OTF$ beamformer, indicated by the five ring resonators of subsystem 3 in Figure~\ref{fig:OBFNschematic}. As mentioned earlier, manual tuning of these five rings has already been shown to achieve the desired delay values~\cite{orr1}, and automatically tuning these five rings should give no problems. However, tuning these five rings for $24$ different beamformer paths using only one scalar value as feedback is much more challenging, as this would increase the number of variables to be tuned simultaneously to $240$. Still, this number is a drastic reduction compared to the situation where all heaters in the full photonic beamformer are tuned simultaneously ($1440$ variables). The DONE algorithm has already been successfully applied to a problem with $150$ variables in the past~\cite{DONEpaper}, so increasing this number to $240$ is not unachievable in the near future.

\section{Conclusion}

We have proposed a novel photonic beamformer for a transmit phased array antenna.
The beamformer is based on optical ring resonators and is fully integrated on a chip.
We have investigated manual and automatic procedures for tuning the delays of  the photonic beamformer.
Automatic procedures are essential in real-life applications where thousands of actuators are considered and where a limited number of variables can be measured, such as the signal-to-noise ratio.
Both the manual and automatic procedures provided beamforming functionalities with a phase error of less than $11.25\degree$ over the whole frequency band that contains signal information.
This is the first time that part of this transmit antenna tile has been tuned automatically.

The system is designed with aircraft-satellite communication in mind as the main application, providing satellite Internet connections on board the aircraft using the $\Ku$-band.
The separate carrier tuning principle greatly reduces the operating bandwidth for the photonic beamformer, from $29$ GHz to $0.5$ GHz.
The ring resonator-based subsystems of the beamformer make squint-free beamforming possible.
A fully automated transmit antenna tile remains for future work.

\appendices

\section{Measurement setup}

\begin{figure*}[tbp]
\centering
\includegraphics[width=0.86\textwidth]{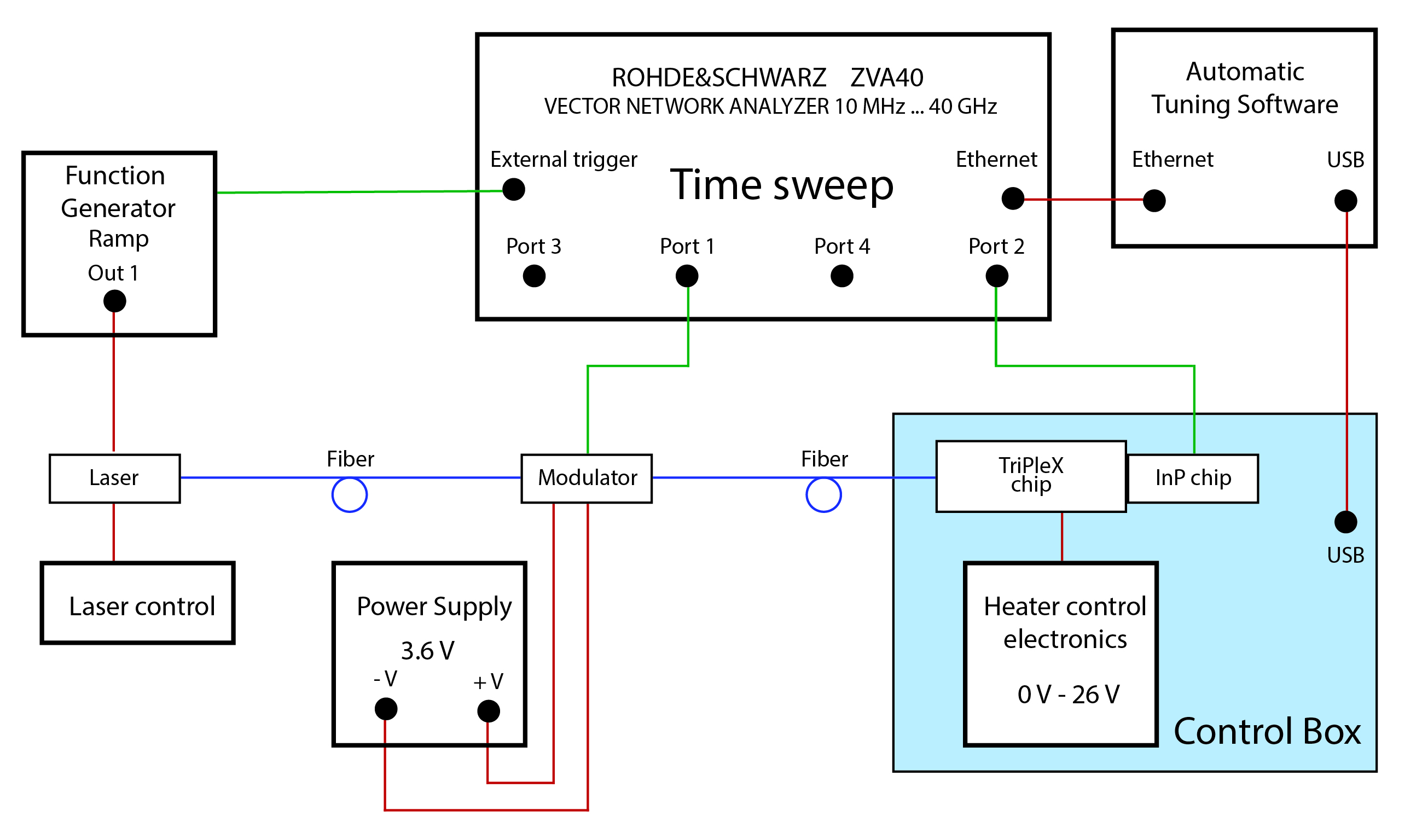}\\%
 \caption{Measurement set-up for the frequency response measurements used in this paper.
}
 \label{fig:timesweep}
\end{figure*}

Figure~\ref{fig:timesweep} shows the measurement set-up for the frequency response measurements shown in this paper.
A Rohde\& Schwarz ZVA40 VNA generated a $50$ MHz RF signal to modulate a laser with a Mach-Zehnder modulator.
The modulated optical signal was coupled into the beamformer control box.
The control box was connected via USB to a laptop with heater control software.
This laptop also contained the automatic tuning software and software for reading the VNA measurements.
The integrated photodetectors inside the control box were connected with RF cables to the second VNA port.
Although integrated modulators were also available, these have not been used in order to prevent crosstalk with the integrated detectors.

The VNA measurements were sent to the laptop with the automatic tuning and heater control software via an ethernet cable.
The phase-shift method~\cite[Sec. IV-A]{novel2} was used for the group delay and power response measurements, using a function generator for the laser.
With this method, the phase response shown on the VNA is actually equivalent to the group delay response of the system, so the objective~\eqref{eq:databased} minimized by the DONE algorithm is actually equal to the mean square error between the desired and measured group delay response.
In Figures~\ref{fig:OBFNtune} and~\ref{fig:SCT}, after the minimization procedure, the phase response was measured directly using a frequency sweep on the VNA (without the external trigger from the function generator).
Here, the laser current was chosen in such a way that the optical carrier would be at 0 GHz in Figure~\ref{fig:OSBFtune} after tuning the OSBF.

\section{Algorithm settings}

The automatic tuning algorithm used in this paper contains several hyper-parameters that are explained and investigated in~\cite{DONEpaper}. Their values as used in this paper are shown in Table~\ref{tab:DONEparam} for the OSBF and OBFN tuning results. The last value indicates the size of the sliding window: only the last $60$ measurements are used in updating the model used by the algorithm, using the same adaptation of the algorithm as in~\cite{pozzi2017high}.

\begin{table}[tb!]
\caption{Values of the hyper-parameters used by the automatic tuning algorithm.
\label{tab:DONEparam}}
\centering
\begin{tabular}{|p{0.3\columnwidth}|p{0.2\columnwidth}|p{0.2\columnwidth}|}
\hline
 Hyper-parameter & Value (OSBF tuning) & Value (OBFN tuning)\\
\hline
Total num. of \mbox{measurements} & $2000$ & $1000$\\
\hline
Num. of basis functions & $3000$ & $3000$\\
\hline
Regularization \mbox{parameter} & $0.1$ & $0.1$\\
\hline
Standard deviation of frequencies & $1.0$ & $1.0$\\
\hline
Exploration parameter & $0.01$ & $0.05$\\
\hline
Sliding window size & $60$ & $60$\\
\hline
\end{tabular}
\end{table}

\bibliographystyle{IEEEtran}
\bibliography{mybib}

\end{document}